\documentclass[%
aps,
 prb,
 amsmath,amssymb,
preprint,%
superscriptaddress,
]{revtex4-1}

\usepackage{graphicx}
\usepackage{dcolumn}
\usepackage{bm}
\usepackage[mathlines]{lineno}

\usepackage[utf8]{inputenc}
\usepackage[T1]{fontenc}
\usepackage{mathptmx}
\usepackage{textgreek}
\usepackage{upgreek}
\usepackage{gensymb}

\draft 

\begin{document}

\title{Combined electrostatic and strain engineering of BiFeO$_3$ thin films at the morphotropic phase boundary }

\author{Johanna Nordlander}
\email[]{jnordlander@fas.harvard.edu}
\affiliation{Department of Materials, ETH Zurich, CH-8093 Zurich, Switzerland}
\affiliation{Department of Physics, Harvard University, 02138 MA, USA}
\author{Bastien F. Grosso}
\affiliation{Department of Materials, ETH Zurich, CH-8093 Zurich, Switzerland}
\author{Marta D. Rossell}
\affiliation{Electron Microscopy Center, Empa, CH-8600 D\"ubendorf, Switzerland}
\author{Aline Maillard}
\author{Elzbieta Gradauskaite}
\author{Nicola A. Spaldin}
\author{Manfred Fiebig}
\author{Morgan Trassin}
\email[]{morgan.trassin@mat.ethz.ch}

\affiliation{Department of Materials, ETH Zurich, CH-8093 Zurich, Switzerland}


\begin{abstract}
Multiferroic BiFeO$_3$ (BFO) possesses a rich phase diagram that allows strain tuning of its properties in thin-film form. In particular, at large compressive strain, a supertetragonal (T) phase with giant polarization is stabilized over the more common rhombohedral (R) structure. To utilize the functionality of such metastable BFO phases in device applications, it is essential to understand the ferroelectric phase evolution upon insertion in nanoscale heterostructures. Here, we explore the emergence of ferroelectric phases close to the morphotropic phase boundary in compressively strained BFO during thin-film growth using in-situ optical second harmonic generation. We find that the epitaxial films form at the growth temperature in the ideal T phase without critical thickness for the polarization. Signatures of T-like and R-like monoclinically distorted phases only appear upon sample cooling. We furthermore demonstrate a robustness of single-domain polarization in the high-temperature T phase during the growth of capacitor-like metal\,|\,ferroelectric\,|\,metal heterostructures. Here, a reduction in tetragonality of the T phase, rather than domain formation, lowers the electrostatic energy. At this lower tetragonality, density-functional calculations and scanning transmission electron microscopy point to the stabilization of a new metastable R-like monoclinic structure upon cooling the heterostructure down to room temperature. Our results thus show that the combination of strain and electrostatic phase stabilization in BFO heterostructures yields a prominent platform for exploring ferroelectric phases and realizing ultrathin ferroelectric devices.
\end{abstract}

\maketitle

\section{Introduction}
Epitaxial strain engineering in complex-oxide thin films has proved to be a successful path for designing materials with novel or enhanced functionality. \cite{Ramesh2007} In the case of ferroelectric oxides, epitaxial strain can lead to enhanced ordering temperatures, different domain configurations and even stabilization of new, metastable phases. \cite{Schlom2007,Damodaran_2016} A prototypical example for the immense impact epitaxial strain can have on ferroelectric properties is BiFeO$_3$ (BFO). In this system, strain engineering led to the discovery of a strain-driven morphotropic phase boundary with a transition from the rhombohedral-like monoclinic phase to a metastable tetragonal-like monoclinic phase at epitaxial compressive strain values exceeding $-$4\%. \cite{zeches2009strain} The epitaxially stabilized tetragonal-like phase has an unusually large tetragonality, corresponding to a ratio between out-of-plane $c$- and in-plane $a$-lattice parameters of $c/a > 1.2$, and a giant spontaneous polarization of $\sim 150\,\upmu \text{C}/\text{cm}^2$ along the $c$ axis. \cite{zhang2011microscopic} Furthermore, because of the flat energy landscape found around morphotropic phase boundaries, \cite{Fu2000,Hatt2010} highly strained BFO films tend to relax their strain state with increasing thickness through the formation of rhombohedral-like monoclinic phase inclusions in the tetragonal-like matrix. In this mixed-phase region of the thickness--strain diagram, the system exhibits exceptionally pronounced electronic, piezoelectric and ferroelectric responses. \cite{zhang2011microscopic,Sharma_2018,Seidel2014electronic}

Despite the promising features displayed by highly strained BFO, it is not yet understood how these metastable polar phases evolve in the technologically relevant ultrathin regime. Such insight is essential to expedite their device implementation. \cite{Yamada2013} Since BFO is polar at the epitaxial growth temperature, \cite{DeLuca2017a} it is of particular importance to understand the interplay between strain and depolarizing-field effects on the formation of the ferroelectric state right in the growth environment in both single layers and device-like heterostructures.

Here, we use Ce$_{0.04}$Ca$_{0.96}$MnO$_3$ (CCMO-) buffered (001)-oriented LaAlO$_3$ (LAO) as metallic substrate for the BFO thin-film growth. The CCMO film is fully strained to the LAO substrate and thus maintains the same $-4.5\%$ lattice mismatch to BFO as LAO. This heterostructure allows us to stabilize the metastable T-like phase of BFO close to the morphotropic phase boundary,\cite{zeches2009strain} yet here with a conducting buffer layer. We use in-situ optical second harmonic generation (ISHG) during growth to probe the emergence of polarization. \cite{DeLuca2017a, Sarott2021} At the deposition temperature, we find that the compressive strain imposed by the substrate results in the epitaxial growth of ideal T-phase BFO in a single-domain state with zero critical thickness for a thickness of up to at least 80 unit cells. Signatures of monoclinic distortion of the T phase and formation of rhombohedral-like monoclinic phase inclusions are only observed to occur upon sample cooling at ca.\ 460\degree C and 200\degree C, respectively. Through the epitaxial design of a CCMO\,|\,BFO\,|\,CCMO capacitor heterostructure, we reveal the robustness of the single domain polarization state of the high-temperature phase. Upon capping ultrathin BFO with a top electrode, the single domain state is preserved and we detect a lowered tetragonality rather than a complete suppression of net polarization as response to incomplete screening of the depolarizing field. Density-functional calculations, in combination with scanning transmission electron microscopy (STEM), show how this reduction in tetragonality further triggers a phase transition to a previously unknown monoclinic phase upon sample cooling toward room temperature. The robustness of polarization in the ultrathin regime of compressively strained BFO, mediated by a metastable polar phase as demonstrated in this work, offers a route to avoid uncontrolled domain formation in favor of a deterministic single-domain polarization state  -- a key ingredient for the design of nanoscale ferroelectric devices.

\section{Results}

The BFO films are grown by pulsed laser deposition on LAO substrates with and without a conducting CCMO buffer layer. The CCMO layer is grown at a substrate temperature of 700\degree C with an energy fluence of 1.15\,J/cm$^2$, whereas BFO is deposited at 670\degree C with 1.3\,J/cm$^2$ fluence. Both layers are grown at 0.15 mbar oxygen partial pressure. The CCMO layers are kept at a 15-unit-cell thickness so as to maintain the in-plane lattice parameter of the LAO substrate, whereas the BFO layer thickness is varied from 10 to 80 unit cells. (In the following, both CCMO and BFO are referred to in terms of their pseudocubic single-formula unit cells.) The thicknesses of the layers are controlled using in-situ reflection high-energy electron diffraction (RHEED) and post-deposition x-ray reflectivity. Reciprocal space mapping by x-ray diffraction is used to characterize the phase composition and orientation of the thin films.

To probe the emergence of polarization in highly strained BFO, we monitor the ISHG response of the films during deposition in a reflection measurement geometry as described in Ref. \onlinecite{DeLuca2017a}. SHG is a symmetry-sensitive nonlinear optical process describing the emission of frequency-doubled light. This process is parametrized by the second-order nonlinear susceptibility $\chi^{(2)}$ and depends on the point-group symmetry of the material. SHG is in particular sensitive to symmetry breaking resulting from the onset or change of direction of spontaneous polarization in a material. \cite{Fiebig2005,Denev2011a,Nordlander2018} We take advantage of this symmetry and polarization sensitivity of SHG to monitor the polar state of our films (see Methods). SHG has previously proved effective in distinguishing the different polar phases in strained BFO films. \cite{Kumar2010,Haislmaier2013,Trassin2015} By combining the SHG probe with the thin-film deposition process, the in-situ nature of our technique now allows us to access directly the spontaneous polarization in BFO as it evolves during growth and also to follow the temperature-dependent phase transitions BFO experiences during post-deposition sample cooling in the growth chamber.

\begin{figure}
    \includegraphics[scale = 0.6]{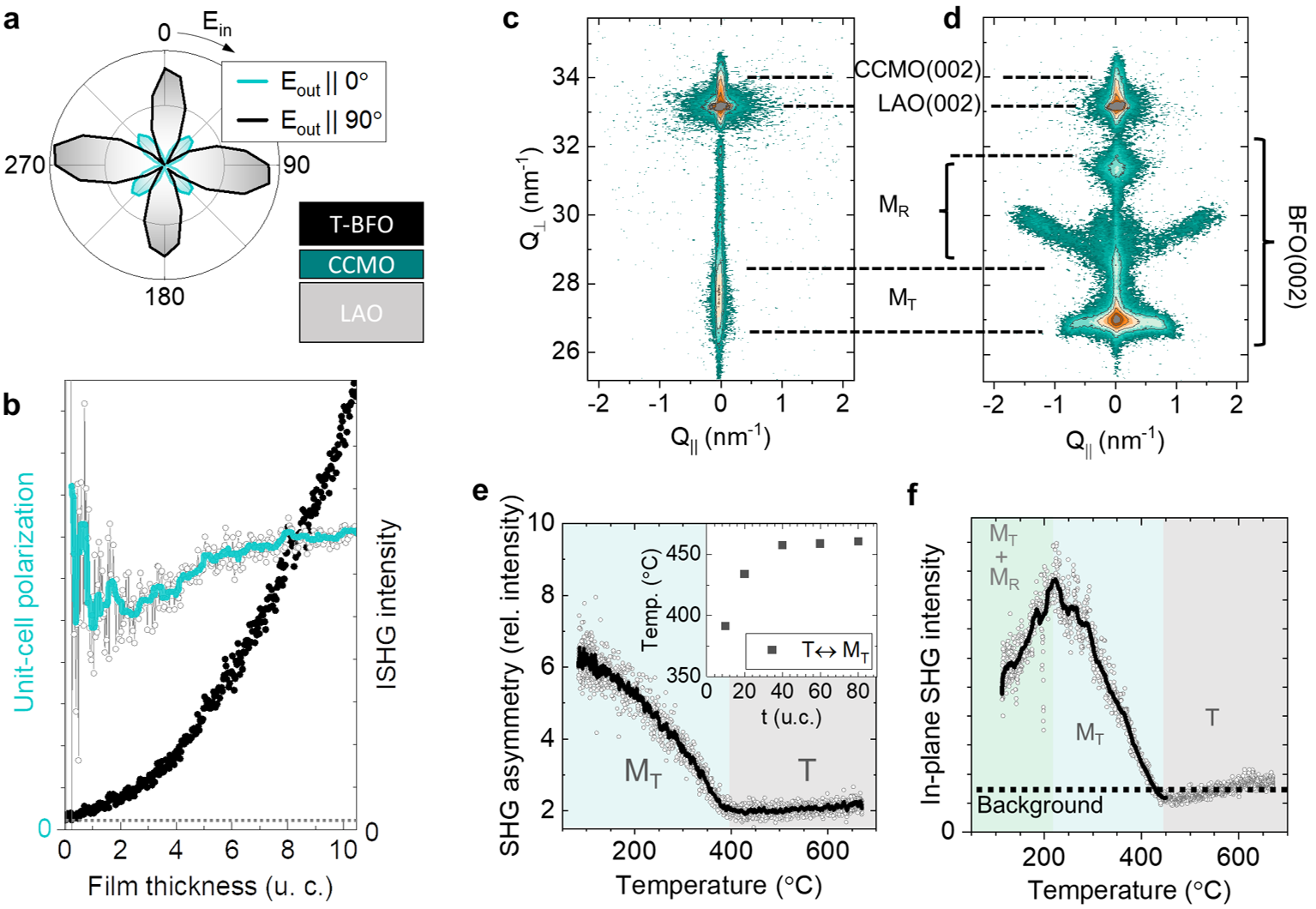}
    \caption{\label{fig:crit_thickness} (a) Polarimetry of the ISHG signal from BFO measured in reflection at the growth temperature. The four-fold symmetry is compatible with a 4$mm$ tetragonal ferroelectric state, and remains unchanged for film thicknesses up to at least 80 unit cells (u.c.), the limit of our experiments. Here, light polarization E$_{\mathrm{in,out}}$ along 0\degree\ (90\degree) is perpendicular (parallel) to the plane of reflection.  (b) The ISHG signal detected at E$_\text{in}$ || E$_\text{out}$ || $90\degree$ during growth (black) and the extracted polarization of the film normalized by the film thickness (gray). The dotted horizontal line represents the background SHG level. A 20-point floating average (blue), to reduce artifacts from normalization of zero-signal noise, highlights the onset of the finite polarization directly at the start of the deposition, hence yielding a zero critical thickness. (c, d) Room-temperature x-ray reciprocal space mapping around the (002) reflections of BFO films grown on CCMO-buffered LAO substrates with thicknesses of (c) 10 u.c.\ and (d) 80 u.c. The ultrathin films remain in the M$_{\text{T}}$ phase, whereas strain relaxation in thicker films promotes phase coexistence between several M$_{\text{T}}$ and M$_{\text{R}}$ phases. (e, f) Temperature dependence of the ISHG response during post-deposition cooling. (e) The ratio between SHG signals detected at E$_\text{in}$ || E$_\text{out}$ || $90\degree$ and E$_\text{in}$ $\perp$ E$_\text{out}$ || $90\degree$ (as defined in (a)) for a BFO film with a thickness of 10 u.c.\ on CCMO-buffered LAO reveals a change of slope with temperature at $390\degree$ C, where the onset of monoclinic distortion is expected. The transition temperature increases with film thickness and saturates around $460\degree$ C (see inset). (f) For a BFO film for which room-temperature characterization reveals the presence of the M$_{\text{R}}$ phase (here, 70 u.c. BFO on LAO), an additional symmetry change can be observed around $200\degree$ C in the SHG signal polarized in the plane of the film (E$_\text{in}$ || E$_\text{out}$ || $0\degree$).}
\end{figure}

We first investigate the emergence of spontaneous polarization in real time during the thin-film growth. Figure~\ref{fig:crit_thickness}(a,b) displays the onset of ISHG, and, hence, spontaneous polarization, when BFO is deposited on CCMO-buffered LAO. Polarimetry of the ISHG signal [Fig.~\ref{fig:crit_thickness}(a)] is compatible with a tetragonal point group for our BFO films. This observation is in agreement with the high-temperature T phase without monoclinic distortions reported previously.\cite{Beekman2013} We find that all our films of this type, in a thickness range up to at least 80 unit cells (the largest value investigated here), grow in this T phase. In other words, we do not observe a symmetry-changing phase transition as a function of thickness at the growth temperature.

Close inspection of the ISHG response at the early growth stages reveals an onset of SHG from the deposition of the very first monolayer. By normalizing the ISHG intensity to the film thickness, we extract the evolution of the spontaneous unit-cell polarization of the film, P$_\text{S} \propto \chi^{(2)}$, \cite{DeLuca2017a} during growth [Fig.~\ref{fig:crit_thickness}(b)]. This confirms that T-phase BFO grown on CCMO-buffered LAO exhibits a spontaneous polarization from the very first unit-cell layer, and the size of the polarization remains roughly independent of thickness. This is noteworthy because conventionally the polarization discontinuities at the top and bottom interfaces of a thin ferroelectric layer lead to a strong depolarizing field due to incomplete screening of bound charges. \cite{Junquera2003,Tagantsev2006,Jia2007} Consequently, ferroelectrics often exhibit a critical thickness below which the spontaneous polarization is either completely suppressed \cite{Fong1650,Junquera2003} or coerced into a nanoscale multidomain state, \cite{Lichtensteiger2005,Lichtensteiger2007,Strkalj2019} such that the net polarization is quenched. In particular, rhombohedral-like monoclinic BFO films under moderate in-plane strain of about -1\% exhibit a critical thickness of five unit cells followed by a thickness-dependent value of P$_\text{S}$ at the growth temperature.\cite{DeLuca2017a} Here, in contrast, we demonstrate a complete absence of critical thickness with an immediate onset of the nearly thickness-independent polarization in the epitaxially stabilized tetragonal phase on a conducting buffer during deposition. 

We further note that the threshold thickness for the emergence of spontaneous polarization is only two unit cells even when the bottom electrode is omitted (see Supplementary Note 1). Such a critical thickness of less than 1\,nm in the absence of metallic depolarizing-field screening further highlights the exceptional robustness of the polarization state in purely tetragonal BFO.

The in-plane lattice parameter of 3.79\,\AA\ imposed on the films by the substrate places BFO close to the morphotropic phase boundary. Hence, whereas at the elevated growth temperature the purely tetragonal phase of the BFO films prevails, we expect a monoclinic distortion of this T phase to take place together with a development of phase coexistence with rhombohedral-like monoclinic phase inclusions towards room temperature.\cite{zeches2009strain} To follow the temperature-dependent evolution of the polarization in this compressively strained BFO system, we investigate the ISHG response for different film thicknesses during post-deposition cooling in the growth chamber. Since the onset of a monoclinic distortion corresponds to a reduction of the point-group symmetry from 4$mm$ in the tetragonal phase to $m$ in the monoclinic phase, such transitions are accompanied by new as well as modified components in the $\chi^{(2)}$ tensor. Therefore a corresponding change in the SHG contributions from the BFO film is expected.\cite{Kumar2010,Haislmaier2013, Trassin2015,Muller2021training} Because of the complexity of the phase diagram at the morphotropic phase boundary in BFO, with several coexisting monoclinic phases reported in the literature, we will henceforth restrict ourselves to a simplified notation. Here, as indicated in Fig.~\ref{fig:crit_thickness}(c, d), a monoclinically distorted phase that preserves the lattice parameters close to that of the parent T phase [Fig.~\ref{fig:crit_thickness}(c)] will be denoted collectively as M$_{\text{T}}$ phases (in the literature also denoted by M$_{\text{A}}$/M$_{\text{C}}$, \cite{Christen2011} M$_{\text{II}}$,\cite{Damodaran2011} or T',\cite{Beekman2013} among others). We will further denote the monoclinic phases that tend to form as inclusions in the M$_{\text{T}}$ matrix,\cite{zeches2009strain,Damodaran2011} and have lattice parameters approaching those of the unstrained rhombohedral phase, as M$_{\text{R}}$ phases. The latter notation therefore encompasses the phases outlined in  Fig.~\ref{fig:crit_thickness}(d) and which have been denoted S', M$_{\text{I}}$, and R' in the literature.\cite{Damodaran2011,Beekman2013} 

By tracking different $\chi^{(2)}$ tensor components in our temperature-dependent ISHG measurements, we identify two transition temperatures upon post-deposition cooling [Fig.~\ref{fig:crit_thickness}(e,f)]. The first transition is observed in the asymmetric temperature evolution of SHG components that are allowed in both the tetragonal 4$mm$ and monoclinic $m$ point groups, with onsets in the range of 460\degree C down to 390\degree C, depending on thickness [Fig.~\ref{fig:crit_thickness}(e, inset)]. The second transition occurs around 200\degree C and is here observed in the SHG component that couples to in-plane polarized states and thus not allowed in the tetragonal 4$mm$ high-temperature phase [see Fig.~\ref{fig:crit_thickness}(f)]. With the excellent agreement of these transition temperatures with values reported in the literature from scanning-probe and x-ray diffraction studies, \cite{Liu2012,Damodaran2012,Beekman2013} we can assign the high-temperature transition to most likely be the point at which the strain-stabilized, purely tetragonal phase develops a monoclinic tilt, i.e., BFO enters an M$_{\text{T}}$ phase, see Fig.~\ref{fig:crit_thickness}(e). The second transition then represents the formation of M$_{\text{R}}$ phase inclusions in this M$_{\text{T}}$ matrix, leading to a M$_{\text{R}}+$M$_{\text{T}}$ phase coexistence. The transition from the T to the M$_{\text{T}}$ phase is seen in all films. In contrast, signatures of the low-temperature transition [Fig.~\ref{fig:crit_thickness}(f)], likely related to the emergence of M$_{\text{R}}$ inclusions, can only be observed if the BFO film thickness exceeds approximately 30-40 u.c. At this thickness, we also start seeing the presence of M$_{\text{R}}$ inclusions at room temperature by atomic force microscopy (Supplementary Fig. S2) and x-ray reciprocal space mapping [Fig.~\ref{fig:crit_thickness}(d)]. The fact that the two transitions only occur below the deposition temperature serves as additional confirmation that the films are grown in the purely tetragonal phase without a monoclinic distortion and remain coherently strained to the substrate throughout the growth process. 

\begin{figure}
    \includegraphics[scale = 0.75]{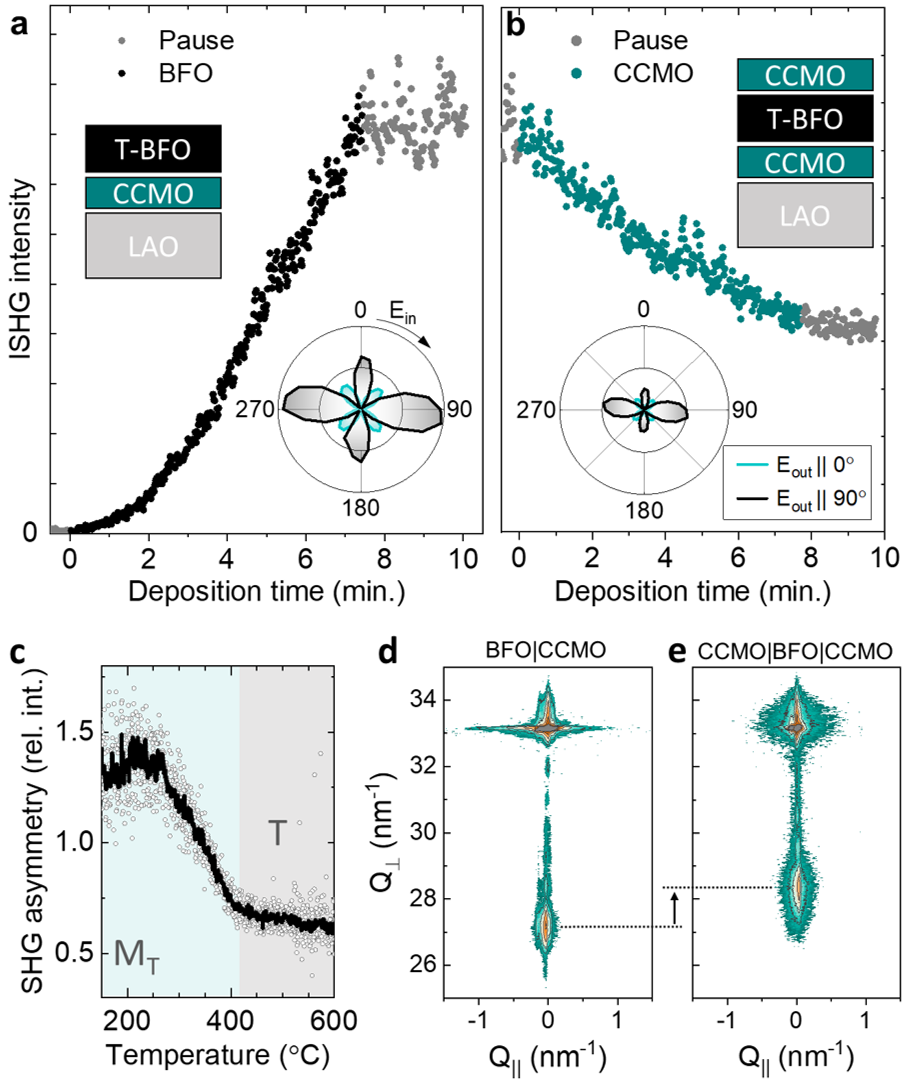} 
    \caption{\label{fig:capping} (a, b) ISHG measurement of polarization in tetragonal BFO during deposition of a multilayer system. (a) Deposition of the BFO film on a CCMO buffer yields zero critical thickness and deposition in the tetragonal phase. The film thickness is 20 u.c. (b) A 15-u.c.\ CCMO capping layer on top of the BFO film leads to a gradual reduction in ISHG intensity. The insets show ISHG polarimetry plots measured as in Fig. \ref{fig:crit_thickness} after the respective layer deposition. The films retain their symmetry after CCMO deposition and only an overall reduction in ISHG magnitude is observed. (c) Tracking of SHG asymmetry in terms of the relative intensity of SHG components (E$_\text{in}$ $\perp$ E$_\text{out}$ || $90\degree$ and E$_\text{in}$ || E$_\text{out}$ || $90\degree$)  during cooling of the capped BFO sample indicates a post-deposition M$_\text{T}$ phase transition at similar, albeit slightly lower, temperature as uncapped BFO films of the same thickness [Fig. \ref{fig:crit_thickness}(e)]. (d,e) X-ray reciprocal space maps of the out-of-plane (002) reflections for a 20-u.c.\ BFO on CCMO-buffered LAO, (d) without and (e) with a CCMO top layer. From the shift of the BFO(002) reflection toward a larger Q value in the film after capping (indicated by the arrow), we extract a reduction of the $c$ lattice parameter of BFO of about 4\%, and correspondingly a reduced tetragonality ($c/a-1$) of about 22\%.}
\end{figure}

Next, we investigate the stability of the high-temperature tetragonal phase in a CCMO|BFO|CCMO capacitor-like heterostructure. In general, a capping layer on top of an ultrathin ferroelectric film impacts the electrostatic boundary conditions at the created interface \cite{Junquera2003} and may induce a transient reduction of the screening of the depolarization field. This can for example stem from a reduced charge-screening efficiency provided by very thin metallic layers (as is the case at the early stages of top electrode deposition), as compared to that provided by the ionic species generally present in the PLD growth environment.\cite{Strkalj2019} As a result, a loss of net polarization in the capacitor can occur, even if the ferroelectric layer was initially grown in a single-domain state. Most importantly, the strain-driven morphotropic phase boundary of BFO is, in addition, highly electric-field sensitive and such changes in the electrostatic conditions may trigger M$_{\text{T}}$ to M$_{\text{R}}$ phase transitions. Here, our ISHG probe brings unique insight into such transient effects during the capacitor growth. \cite{Strkalj2019,Strkalj2020,Sarott2020,Nordlander2021}

In Fig.~\ref{fig:capping}(a,b), we show the ISHG response during growth of a 20-unit-cell tetragonal BFO film and the subsequent deposition of a 15-u.c.\ CCMO capping layer, which results in the capacitor architecture. During deposition of the CCMO top layer, the ISHG intensity related to the single-domain polarization in the tetragonal BFO films exhibits a continuous reduction down to about 50\% of its initial value. This decrease cannot be solely attributed to optical losses in the CCMO layer, which we estimate result in at most a 33\% reduction of the ISHG yield (see Supplementary Note 3). It is also incompatible with an immediate quench of the net polarization caused by a depolarizing-field-induced domain splitting, which would yield a total suppression of the ISHG signal. \cite{Strkalj2019} The unperturbed tetragonal symmetry throughout the growth of the CCMO electrode, confirmed by the ISHG polarimetry in Fig.~\ref{fig:capping}(a, b), further excludes the nucleation of M$_\text{R}$ phase inclusions with differently oriented polarization. In fact, analogous to the uncapped films discussed above, the onset of monoclinic distortions is only observed when cooling the sample [Fig.~\ref{fig:capping}(c)]. As a hint towards the origin of the loss of SHG intensity, post-deposition x-ray reciprocal space mapping on films before and after CCMO capping reveals that the top electrode deposition results solely in a reduction of the BFO $c/a$ ratio, from ca.\ 1.22 to 1.17. The corresponding reduction of the tetragonality $c/a-1$, i.e. where zero tetragonality indicates a cubic phase with $c=a$, is about 22\% [Fig.~\ref{fig:capping}(d,e) and Supplementary Fig. S3]. Because of the intimate relation between tetragonality and polarization, we attribute this decrease in the $c$ lattice parameter to a polarization reduction in the BFO layer during capping, caused by a residual influence of the depolarizing field, commonly observed in rather thick ferroelectric layers.\cite{Lichtensteiger2005} Indeed, as has been reported previously,\cite{Yamada2013ferroelectric} the electrical conductivity, and thus the charge-screening efficiency, of CCMO films tends to decrease drastically at very low film thicknesses, where a dead (insulating) layer thickness of 5 u.c.\ also was suggested. As such, the reduced polarization would account for that part of the reduction of ISHG intensity that is not associated with optical loss. Remarkably, we still observe a non-zero ISHG intensity after the deposition of the CCMO capping layer. This means that a net polarization still remains in the heterostructure despite the ultralow thickness of 20 u.c.\ of the BFO film, where domain splitting or complete polarization suppression would be expected in the presence of a partially unscreened depolarizing field. \cite{Strkalj2019}

\begin{figure*}
    \includegraphics[width=\textwidth]{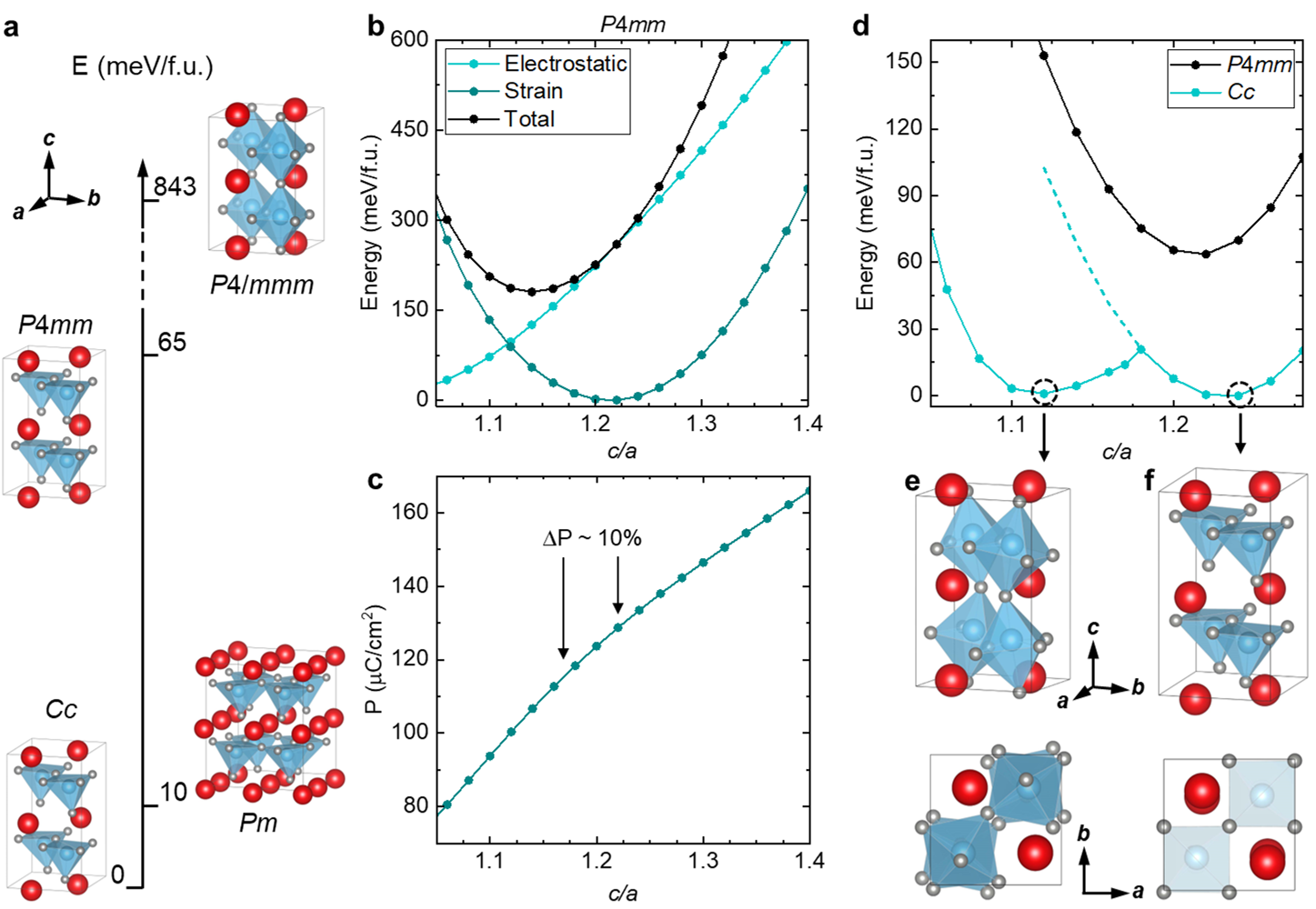} 
    \caption{\label{fig:theory}(a) Calculated DFT structures and relative energies of the \textit{P}4/\textit{mmm}, \textit{P}4\textit{mm}, \textit{Pm} and \textit{Cc} phases on a LAO substrate. (b). Energy of the \textit{P}4\textit{mm} phase calculated for different $c/a$ ratios including the depolarizing field effects (Total) and decomposed into purely strain energy (Strain), calculated with DFT, and electrostatic energy (Electrostatic), computed using the electrostatic model presented in Ref. \onlinecite{Mundy2022liberating}. A non-zero depolarizing field favours a lower tetragonality (minimum of the Total curve) in contrast to a larger tetragonality at zero depolarizing field (minimum of Strain curve). (c) Calculated polarization of the \textit{P}4\textit{mm} phase of BFO for different $c/a$ ratios at fixed $a$. The arrows indicate the experimentally determined reduction of tetragonality due to CCMO capping of ultrathin BFO. The corresponding reduction in P for the \textit{P}4\textit{mm} phase is ca. 10\%. (d) Strain energy for the \textit{P}4\textit{mm} and \textit{Cc} phases as a function of $c/a$. Two metastable monoclinic phases are found at both minima of the light blue curve. (e) Schematic of the \textit{Cc} structure obtained at a $c/a$ ratio of 1.12 along two orientations. (f) Schematic of the \textit{Cc} structure obtained at a $c/a$ ratio of 1.24 along two orientations. Note that the $Cc$ phase at larger $c/a$ ratio is the M$_\text{T}$ phase, which at smaller $c/a$ ratio (below 1.18) follows the dashed line curve in (d), thus becoming higher in energy. Nevertheless, we discover a second phase having its energy minimum (less than 1 meV/f.u. above that of the M$_\text{T}$ phase) at smaller $c/a$ ratio, and therefore energetically more favorable than the M$_\text{T}$ phase for $c/a$ ratios below 1.18.}
\end{figure*}

To shed light on the polar state of our capped BFO films and to understand the mechanism that supports the robustness of polarization in the ultrathin layer in spite of residual depolarizing field effects, we turn to density-functional theory (DFT) calculations. In Fig.~\ref{fig:theory}(a), we show our calculated structures for compressively strained bulk BFO, together with their relative energies, for the following phases (and corresponding space groups): monoclinic M$_\text{T}$ (\textit{Cc} and \textit{Pm}), tetragonal T (\textit{P}4\textit{mm}) and tetragonal high-symmetry centrosymmetric (\textit{P}4/\textit{mmm}). To match our experimental observations, the BFO in-plane lattice constant $a$ is kept fixed for all structures to that of the LAO substrate, which we calculate to be 3.77\,\AA. 

Consistent with earlier calculations,\cite{Christen2011} we find that the lowest-energy monoclinic structure belongs to the \textit{Cc} phase rather than the \textit{Pm} phase. We therefore choose the \textit{Cc} structure to represent the monoclinic M$_\text{T}$ phase in our analysis. In order to assess the robustness of the polarization in the different phases, we begin by comparing the relative energies between this monoclinic phase and the higher-energy \textit{P}4\textit{mm} T- and \textit{P}4/\textit{mmm} non-polar phases, respectively. As seen in Fig.~\ref{fig:theory}(a), a significantly larger energy is ascribed to the suppression of the polar distortions, yielding the non-polar \textit{P}4/\textit{mmm} phase, than to the removal of the monoclinic distortion from the oxygen-polyhedra rotations (see also Supplementary note 5). This difference in energy scales for the two types of distortions is consistent with earlier experiments, which showed that while oxygen-polyhedra rotations are lost towards higher temperatures, compressively strained BFO retains its polar distortions and thus exhibits a high ferroelectric T$_{\mathrm{C}}$ \cite{Beekman2013}. Hence, we choose the polar \textit{P}4\textit{mm} phase to represent the structure at the growth temperature in our analysis.

We next address the persistence of polarization in this tetragonal phase against domain formation in the ultrathin limit. We obtain a large polarization of $\sim$120\,$\upmu$C/cm$^2$ for T-phase BFO from summing over the displacement of each atom from its reference non-polar \textit{P}4/\textit{mmm} phase multiplied by its Born Effective charges. While $\sim$50\,$\upmu$C/cm$^2$ is compensated by the discontinuity in layer polarization at the BFO|CCMO interface (since BFO has charged planes while CCMO does not\cite{Spaldin2021Layer}), this still results in a high electrostatic cost for the single-domain configuration in a thin-film sample according to a simple electrostatic model.\cite{Mundy2022liberating} In particular, we find that while in bulk, the equilibrium $c/a$ ratio sits around 1.22, it is shifted towards a reduced $c/a$ ratio when electrostatic boundary conditions are taken into account by introducing an energy cost, as developed in Ref. \onlinecite{Mundy2022liberating}, for having a polar discontinuity at the interface between BFO and its surrounding  [Fig.~\ref{fig:theory}(b)]. Using the reduced $c/a$ value measured by X-ray diffraction on the capped BFO sample displayed in Fig. \ref{fig:capping} as input in our DFT calculations, we find that the corresponding polarization value is about 10\% lower than the polarization at the equilibrium $c/a$ ratio [Fig.~\ref{fig:theory}(c)]. This calculated drop in polarization agrees exceptionally well with the estimated polarization reduction needed to account for the observed drop in ISHG intensity during CCMO capping in addition to optical absorption (see Supplementary note 3). Therefore, we can conclude that the observed reduced tetragonality in the capped BFO layer (as seen in Fig.~\ref{fig:capping}), leading to a reduction of both the net polarization and the electrostatic energy density, is likely a response to the incomplete screening of the depolarizing field by the top CCMO layer. The partial polarization suppression we detect in our tetragonal BFO heterostructures seems to sufficiently lower the electrostatic energy of the system such that domain formation is not needed and, thus, the single-domain state prevails in the ultrathin limit.

Let us now see what happens as we relax the compressively strained BFO system from the high-temperature \textit{P}4\textit{mm} phase by allowing monoclinic distortions to emerge for a range of $c/a$ ratios. As seen in Fig.~\ref{fig:theory}(d-f), we find two distinct energy minima that each could represent our room temperature monoclinic phase. While the $Cc$ phase at $c/a=1.24$ [Fig.~\ref{fig:theory}(f)] is the M$_\text{T}$ phase discussed earlier and matches well with our uncapped BFO films, there is a metastable $Cc$ phase at $c/a=1.12$ [Fig.~\ref{fig:theory}(e)]. Notably, this metastable phase is extremely close in energy to the M$_\text{T}$ phase (the energy difference is less than 1 meV/f.u.), yet it has a lower $c/a$ value close to that found in our capped ultrathin BFO film. This is thus likely the structure we also find experimentally. Indeed, maintaining the M$_\text{T}$ phase at $c/a$ values below 1.18 would be associated with a large energy cost [indicated by the dashed line in Fig. \ref{fig:theory}(d)]. The new metastable phase identified here is still of \textit{Cc} symmetry but possesses the tilt pattern of M$_\text{R}$ (rather than M$_\text{T}$) BFO. We thus label this phase M$_\text{R}^{*}$ as an R-like BFO phase with enhanced (close to T-like) $c/a$ ratio.  As detailed in Table \ref{tab:DFTpolar}, we find that the M$_\text{R}^{*}$ phase has a lower out-of-plane polarization component of about 40\%, while doubling its polarization component along the $a$ axis, as compared to the M$_\text{T}$ phase of uncapped BFO films. Interestingly, this metastable M$_\text{R}^{*}$ phase has a similarly reduced $c/a$ ratio as that calculated for T-phase BFO in the presence of an unscreened depolarizing field ($c/a = 1.14$). These calculations thus suggest that a depolarizing-field-induced drop in tetragonality in the high-temperature T phase of BFO during heterostructure growth may in turn lead to the stabilization of this new monoclinic phase at lower temperature.

\begin{table}[]
    \centering
    \begin{tabular}{|m{2cm} |m{1.7cm} || m{1.5cm} | m{1.5cm} | m{1.5cm}|}
    \hline
      $Cc$ phase & $c/a$ ratio & $P_a$ & $P_b$ & $P_c$ \\
      \hline \hline
      M$_\text{R}^{*}$ &1.12 & 61.7 & 0.6 & 70.2 \\
      \hline
      M$_\text{T}$ & 1.24 & 34 & 0.1 & 128.1 \\
      \hline
    \end{tabular}
    \caption{Calculated polarization components (in $\upmu$C/cm$^2$) in the M$_\text{T}$ and M$_\text{R}^{*}$ monoclinic phases.}
    \label{tab:DFTpolar}
\end{table}

\begin{figure*}
    \includegraphics[scale=0.72]{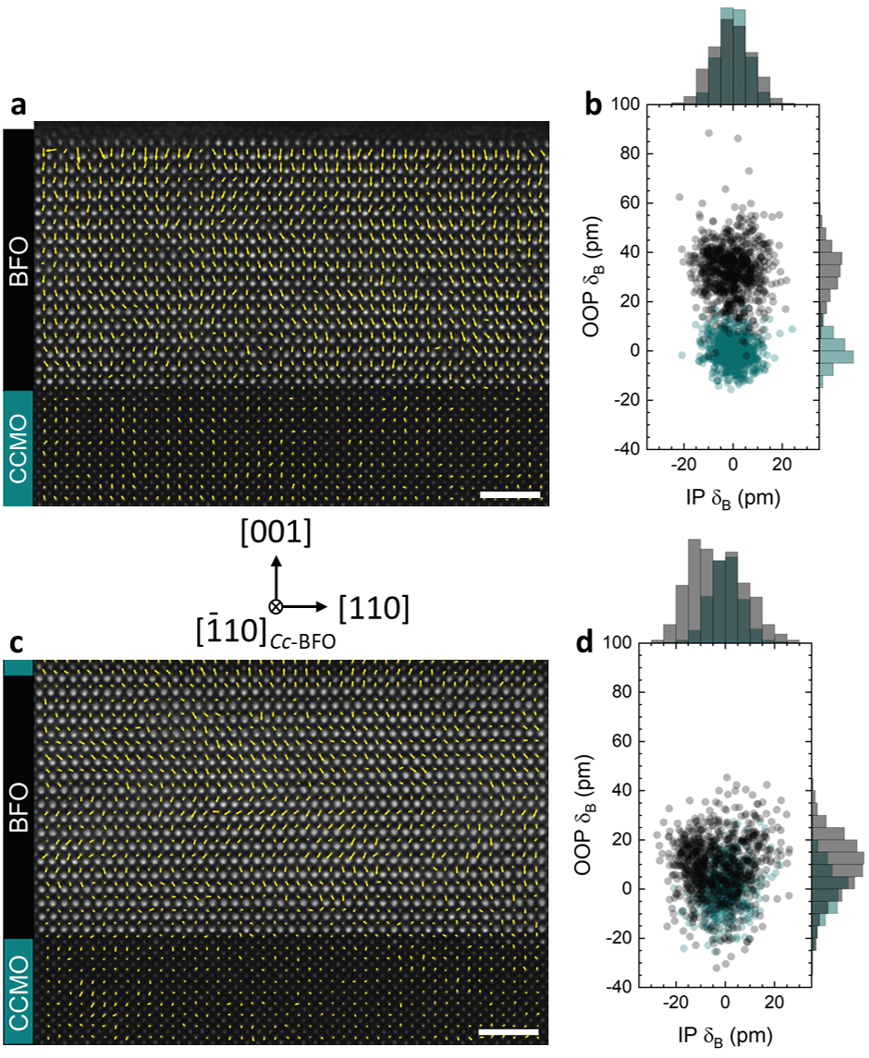}
    \caption{\label{fig:stem} Cross-sectional HAADF-STEM images along the monoclinic [$\bar{1}10$]$_{\mathrm{BFO}}$ zone axis of nominally 20-u.c.-thick BFO films in absence (a) and presence (b) of a CCMO capping layer. (a) Mapping the displacement pattern of Fe$^{3+}$ ($B$-site) ions with respect to the Bi$^{3+}$ ($A$-site) ions in the uncapped BFO yields an average displacement of 32 pm along the out-of-plane [001]$_{\mathrm{BFO}}$ direction and a spread of less than 10 pm (fully overlapping with the CCMO buffer layer) along the in-plane [110]$_{\mathrm{BFO}}$ direction. The ionic displacement amplitudes agree qualitatively with the DFT-calculated values for the M$_{\mathrm{T}}$ phase. (d) In the case of the CCMO capped film of similar thickness as in (a), the out-of-plane Fe$^{3+}$ displacement is suppressed to about 10 pm, while the spread of the in-plane displacement has almost doubled. The scale bars in (a) and (c) are 2 nm}
\end{figure*}

To confirm the experimental stabilization of the M$_\text{R}^{*}$ phase in our capped ultrathin BFO films, we use high-resolution high-angle annular dark-field (HAADF) STEM to map the polar displacement on the Fe site at room temperature (Fig.~\ref{fig:stem}). Compared to the uncapped film [Fig.~\ref{fig:stem}(a)], we note a drop as large as 70\% in the displacement amplitude along the $c$-axis in the capped film [Fig.~\ref{fig:stem}(b)]. In this case, a larger displacement in the $ab$ plane is additionally detected. This polarization state in the CCMO-capped BFO film is in qualitative agreement with the DFT-calculated metastable M$_\text{R}^{*}$ monoclinic phase. It is furthermore in clear contrast to the polar displacement pattern of the M$_\text{T}$ phase seen in the uncapped BFO film of similar thickness [Fig.~\ref{fig:stem}(a)]. We therefore conclude that the electrostatic boundary conditions accompanying the capping of ultrathin T-BFO films in a capacitor-like structure (reducing the $c/a$ ratio in the T phase) results in the stabilization of a new metastable M$_\text{R}^{*}$ BFO phase whose polar displacement pattern resembles the R-phase BFO, yet with a significantly larger tetragonal distortion.

Notably, the net downwards polarization direction seen by STEM in uncapped films (favored by the CaO-FeO$_2$ interface termination between the bottom CCMO and BFO,\cite{Spaldin2021Layer} see Supplementary note 6) is preserved also in the capped film, without macroscopic domain formation [Fig.~\ref{fig:stem}(b)]. This is in line with our ISHG measurements which also suggest the perseverance of a net polarization in the ultrathin, capped BFO films.

\section{Conclusion}
In summary, we have demonstrated an exceptional robustness of polarization in epitaxially stabilized supertetragonal BFO ultrathin films grown on CCMO-buffered LAO. Only upon sample cooling is the monoclinic phase inhomogeneity characteristic of the strain-driven morphotropic phase boundary in BFO observed. We find that the high-temperature T phase of BFO has zero critical thickness on the conducting buffer layer and grows in a single-domain ferroelectric state. This single-domain state is even preserved under unfavorable electrostatic boundary conditions and the notorious imperfect charge screening at the interfaces at high temperature.\cite{Yamada2013ferroelectric, Junquera2003} Density-functional calculations identify a previously unreported metastable monoclinic phase in compressively strained BFO ultrathin films which we stabilize experimentally through a depolarization-field induced lowering of tetragonality during heterostructure growth.

While further studies are needed to fully characterize the new metastable phase reported here, in light of the recent revelation of the influence of epitaxial strain on the BFO antiferromagnetic spin cycloid configuration,\cite{Haykal2020antiferro} we may expect unconventional multiferroic properties in our CCMO capped BFO films. Our findings thus prove the great potential of compressive strain combined with electrostatic engineering of ultrathin BFO films for implementation of versatile ferroelectric states in nanoelectronic devices such as ferroelectric tunnel junctions \cite{Yamada2013} and ferroelectric field-effect transistors.

\section{Methods}
\subsection{Optical SHG} The ISHG experiments were performed in a 90\degree\ reflection geometry using a fundamental laser wavelength of 1200\,nm with a pulse width of 45\,fs and repetition rate of 1\,kHz.\cite{DeLuca2017a} A pulse energy of 20\,$\upmu$J and a spot size of 250\,$\upmu$m in diameter was incident onto the sample. The ISHG intensity was detected at 600\,nm using a monochromator and a photomultiplier tube. The ISHG polarimetry was performed by rotating the polarization of the incident light, E$_\text{in}$, from 0\degree\ to 360\degree, where 0\degree\ and 90\degree\ correspond to light polarized perpendicular to (s) and within (p) the plane of light reflection, respectively. The corresponding  ISHG intensity was detected for both s-polarized (E$_\text{out}=0\degree$) and p-polarized (E$_\text{out}=90\degree$) SHG configurations.

 In the electric-dipole approximation, SHG is expressed by

\begin{equation}
    P_i(2\omega) = \epsilon_0 \chi^{(2)}_{ijk}E_j(\omega)E_k(\omega),
\end{equation}

where the indices $i, j, k$ each take on coordinates $x$, $y$ and $z$, $P_{i}$ is the component of the generated nonlinear polarization, $\epsilon_0$ stands for the vacuum permittivity, $E_{j,k}$ are the electric-field components of the incident light and $\chi^{(2)}_{ijk}$ represents the material-dependent tensor components of the second-order susceptibility. The SHG intensity scales with the film thickness $t$ as $I_{SHG} \propto |\chi^{(2)}t|^2 $, where the set of non-zero $\chi^{(2)}$ components is determined by the crystallographic point-group symmetry. In a ferroelectric like BFO, the magnitude of these tensor components is proportional to the spontaneous polarization P$_\text{S}$.

\subsection{Structural characterization} X-ray reciprocal space mapping was performed using a Panalytical X’Pert$^3$ MRD four-circle diffractometer at a wavelength of 1.5406\,\AA.

Electron transparent samples for STEM investigations were prepared in cross-section by using a FEI Helios 660 G3 UC focused ion beam (FIB) operated at acceleration voltages of 30 and 5\,kV after deposition of C and Pt protective layers. A FEI Titan Themis operated at 300\,kV and equipped with a probe spherical aberration corrector (DCOR, CEOS) and ChemiSTEM technology was used for HAADF-STEM imaging and energy-dispersive X-ray (EDX) spectroscopy. A probe semi-convergence angle of 18 mrad and an annular semi-detection range of the annular dark-field detector of 66-200\,mrad were used. Averaged HAADF-STEM images were obtained after non-rigid registration of  time series consisting of 12 frames (2048x2048, 1\,$\upmu$s) using the SmartAlign software.\cite{Jones2015} The averaged images were background-corrected, denoised and deconvolved as described in Refs. \onlinecite{Campanini2018,Campanini2020}. Subsequently, the atomic column positions were fitted with picometer precision by means of seven-parameter two-dimensional Gaussians using custom-developed MATLAB scripts based on the method proposed in Ref. \onlinecite{Yankovich2014}. Ferroelectric dipole maps were calculated from the relative displacements of the two cation sublattices present in the BFO and CCMO structures. Thus, the local ferroelectric dipoles were calculated by measuring the polar displacement in the image plane of the $B$-site position from the center of mass of its four nearest $A$-site neighbors. In the ferroelectric dipole maps of Fig. \ref{fig:stem} overlaid on the HAADF-STEM images, the dipole moments are plotted opposite to the $B$-site displacement ($\delta_B$). For the graphs in Fig. \ref{fig:stem}, the mean displacement <$\delta_B$> of the non-polar CCMO buffer layer is set to zero (to correct for minor sample misalignments) and the $\delta_B$ values measured on the BFO film are corrected accordingly in relation to the CCMO reference. 

\subsection{Calculations} Our calculations were performed using density functional theory (DFT) with the projector augmented wave (PAW) method \cite{Blochl1994} as implemented in the Vienna ab initio simulation package (VASP).\cite{Kresse1996} Good convergence was achieved with an 8x8x6 $\Gamma$-centred k-point mesh for a 20-atom unit cell, and an energy cut-off of 750\,eV for the plane-wave basis. We used the PBEsol+U functional,\cite{Perdew1996} with U$=4$\,eV on the Fe d orbitals, and imposed G-type antiferromagnetism in all the calculations. We use the ISODISTORT software \cite{ISODISTORT,Campbell2006} to analyze the structural distortions. The electrostatic energy density at the interface was estimated by the coupling between the polarization and the depolarizing field: $\dfrac{1}{2} \dfrac{(P-P_{\mathrm{L}})^2}{\epsilon \epsilon_0}$, with spontaneous polarization $P = 120$\, $\upmu\mathrm{C}/\mathrm{cm}^2$, layer polarization $P_{\mathrm{L}} = 50$\, $\upmu\mathrm{C}/\mathrm{cm}^2$  and $\epsilon = 55$. \cite{Lu2010}

\begin{acknowledgments}
The authors thank Christian Tzschaschel for fruitful discussions. M.T. acknowledges the Swiss National Science Foundation under Project No. 200021\textunderscore188414. J.N., M.T. and M.F. acknowledge financial support by the EU European Research Council under Advanced Grant Program No. 694955-INSEETO. J.N. acknowledges financial support from the Swiss National Science Foundation under Project No. P2EZP2\textunderscore195686. M.D.R. acknowledges the Swiss National Science Foundation under Project No. 200021\textunderscore175926. B.F.G and N.A.S acknowledge financial support from the Koerber foundation. Computational resources for DFT were provided by ETH Zürich and the Swiss National Supercomputing Centre (CSCS), Project ID No. s889.
\end{acknowledgments}

\bibliography{biblio}

\end{document}